\documentstyle[aps,pra,psfig,epsf,epsfig]{revtex}
\draft
\def\br{{\bf r}}
\def\bk{{\bf k}}

\begin{document}
\title{One-dimensional compression of Bose-Einstein condensates
by laser-induced dipole-dipole interactions}
\author{S. Giovanazzi \and D. O'Dell \and G. Kurizki \\
Weizmann Institute of Science,
76100 Rehovot, Israel }
\date{\today}
\maketitle
\begin{abstract}
We consider a trapped cigar-shaped atomic Bose-Einstein condensate 
irradiated by a single far-off resonance laser polarized along the cigar 
axis. The resulting laser induced dipole-dipole
interactions between the atoms significantly change 
size of the condensate, and can even cause its self-trapping.
\end{abstract}

The ability to alter the interatomic potential
of weakly interacting Bose-Einstein condensates (BECs)
 by applying electromagnetic fields
affords a high degree of control
over the Hamiltonian and thus the possibility to engineer the 
macroscopic properties of such many-body systems \cite{burnett98}.
The experiment of Inouye \textit{et al} \cite{inouye98}
has demonstrated how the s-wave scattering length can be changed 
by magnetic fields via a Feshbach resonance.

We have recently proposed a different avenue: 
the use of off-resonant lasers to induce long-range dipole-dipole
interactions in atomic BECs 
\cite{giovanazzi2001a,odell2000,giovanazzi2001c,giovanazzi2001b}. 
These interactions can drastically modify  a BEC, 
causing its self-trapping \cite{giovanazzi2001a}, 
laser induced self-''gravity'' \cite{odell2000} 
(for certain laser-beam configurations), 
``supersolid'' structures \cite{giovanazzi2001c},
and peculiar excitation spectra \cite{giovanazzi2001b}.
At the same time, off-resonant lasers may undergo ``superradiant''
Rayleigh scattering \cite{inouye99} from the condensate, 
concurrently with collective atomic recoil (CARL) \cite{piovella2001}.
It is important to draw a clear distinction between such effects 
and those of laser-induced dipole-dipole forces.
The purpose of this paper is to suggest a simple geometry 
in which the compression of a trapped condensate
and its possible self-trapping are unambiguously 
caused by laser-induced dipole-dipole interactions, 
whereas superradiant Rayleigh scattering or CARL are absent.

\begin{figure}[htbp]
\begin{center}
\centerline{\epsfig{figure=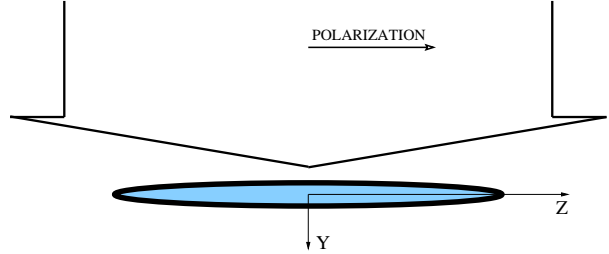,width=8cm}}
\end{center}
\vspace{-4ex}
\caption{The laser beam and condensate geometry. By choosing
the polarization to be along the long axis of the condensate
superradiant effects are suppressed.
}
\label{fig:setup}
\end{figure}

A trapped cigar-shaped BEC that is tightly confined in the radial x-y 
plane is irradiated by a plane-wave laser at a wavelength larger than 
the radial size of the condensate.
The laser polarization is chosen to be along the long z-axis 
of the condensate in order to suppress ``superradiant'' Rayleigh scattering
\cite{inouye99} or CARL \cite{piovella2001} 
that are \emph{forbidden} in the direction of the field polarization.
The interatomic dipole-dipole potential induced by 
far-off resonance electromagnetic radiation of intensity $I$, wave-vector
${\bf q}= q \hat{{\bf y}}$ (along the y-axis), 
and polarization $\hat{{\bf e}} = \hat{{\bf z}}$ (along the z-axis) is 
\cite{thirunamachandran80}
\begin{equation}
U({\bf r}) =
\left(\frac{ {\mathrm I}}{4 \pi c \varepsilon_{0}^{2}}\right)
\alpha^{2}(q) V_{zz}(q, {\bf r})
\cos (q y).   \label{eq:tpot}
\end{equation}
Here ${\bf r}$ is the interatomic axis, $\alpha(q)$ the isotropic,
dynamic, polarizability of the atoms at frequency $cq$, and
$V_{zz}$ is the appropriate component of the 
retarded dipole-dipole interaction tensor
\begin{eqnarray}
V_{zz}=\frac{1}{r^{3}}  \Big[ \big(1 - 3 \cos^2(\theta)\big) 
\big( \cos q r +qr \sin q r \big)
  -  \sin^2(\theta) q^{2}r^{2}
\cos qr \Big]  \label{eq:retarded-dip-int}
\end{eqnarray} 
where $\theta$ is the angle between the  interatomic axis
and the z-axis.
We note that the far-zone ($qr \gg 1$) behavior of the 
dipole-dipole potential (\ref{eq:tpot}) along the z-axis direction
is proportional to $- \sin (qr) / (qr)^2$.

A mean-field description of a BEC with dipole-dipole forces
\cite{odell2000,goral,foldy} can be accomplished through the
Gross-Pitaevskii
equation  \cite{dalfovo99} for the
condensate order parameter
\begin{equation}
i \hbar	 \frac{\partial \Psi }{\partial t} =
\frac{\delta }{\delta \Psi^* } H_{\mathrm{tot}}\;.
\label{ggp}
\end{equation}
The mean-field energy functional
\begin{equation}
H_{\mathrm{tot}} = H_{\mathrm{kin}} + H_{\mathrm{ho}}+
H_{\mathrm{dd}} + H_{\mathrm{s}}
\end{equation}
contains contributions from the one-body energies
and the interatomic potentials:
(a)  the kinetic energy
\begin{equation}
H_{\mathrm{kin}}=\int \, (\hbar^2/2m)
|{\bf \nabla}\Psi|^2  \,d^{3}r \;;
\end{equation}
(b)  harmonic trapping  
\begin{equation}
H_{\mathrm{ho}}=\int
V_{\mathrm{ho}}\,n(\br) \,d^{3}r \;,
\end{equation}
where $n(\br)$ is the atomic number density; 
(c) the dipole-dipole potential
\begin{equation}
H_{\mathrm{dd}} =   (1/2)
\int n(\br) n(\br')
 \, U(\br-\br')\,d^{3}r \, d^{3}r' \;;
\end{equation}
(d) the mean-field energy
due to the very short-range ($r^{-6}$) van der Waals interaction, which will
be treated, as is usual, 
within the delta function pseudo-potential approximation,
\begin{equation}
H_{\mathrm{s}} = (1/2) (4 \pi a \hbar^{2}/m) \int n(\br) d^{3}r\;;
\end{equation}
where $a$ is the s-wave scattering
length.

The integrations involved in the energy functionals 
are simpler in momentum space,
for which $H_{\mathrm{dd}}= (1/2)(2 \pi)^{3/2}$ 
$ \int \widetilde{U}({\mathbf k})
\, \tilde{n}(\bk) \, \tilde{n}(-\bk) \, d^{3}k$.
The Fourier transform of the dipole-dipole potential
(\ref{eq:tpot}),
$\widetilde{U}({\mathbf k})=(2 \pi)^{-3/2} \int  \, \exp [{\mathrm i}
{\mathbf k} \cdot {\mathbf r}] \, U({\mathbf r}) \,  d^{3}r$, can be shown
to be \cite{thirunamachandran80,morice}, for the laser
propagation and polarization as in Fig.~\ref{fig:setup}, 
\begin{eqnarray}
\widetilde{U}({\mathbf k})  = 
\frac{{\mathrm I} \, \alpha^{2}} {2 (2 \pi)^{3/2} \epsilon_{0}^{2} c} 
\left( -\frac{2}{3} 
  +
  \frac{k_{z}^2-q^2}{k_{x}^{2}+(k_{y}-q)^{2}+k_{z}^{2}-q^2}
 + 
 \frac{k_{z}^2-q^2}{k_{x}^{2}+(k_{y}+q)^{2}+k_{z}^{2}-q^2}
\right) \;. 
\end{eqnarray}

We proceed by  adopting a cylindrically symmetric (about
the  axial $\hat{z}$ direction)
variational ansatz for the density. 
For this to hold true requires the tight radial trapping to prevail
over the (anisotropic) dipole-dipole forces in the radial direction.
It is then reasonable to approximate the radial density 
by a Gaussian profile whose width is the variational parameter $w_{r}$:
$n(\br) \equiv N $ $(\pi w_{r}^{2} )^{-1}$ $ n^{z}(z)
\exp \left[-(x^{2}+y^{2})/w_{r}^{2} \right] $, where
N is the total number of atoms and $n^{z}(z)$ is the as yet
unspecified axial density.
Then, in momentum space, the density becomes
\begin{equation}
\tilde{n}(\bk)=  \frac{N}{(2 \pi)} \widetilde{n^{z}}(k_{z})
 \exp \left(-\frac{w_{r}^{2}}{4}(k_{x}^{2}+k_{y}^{2})
\right)\;,
\end{equation}
$\widetilde{n^{z}}(k_{z})$ being the Fourier transform of the
axial density $n^{z}(z)$.
Assuming tight radial confinement with respect to the laser wavelength, 
$\overline{w}_{r} \equiv w_{r}/\lambda_{\mathrm{L}} \ll 1$, 
the radial integration in $H_{\mathrm{dd}}$ 
can be approximately evaluated, so that the dipole-dipole energy
reduces to a one dimensional functional along the axial direction
\begin{eqnarray}
H_{{\mathrm dd}}  =  \frac{1}{2}\int n^{z}(z) 
 n^{z}(z') U^{z}(z-z') \, dz \, dz' 
 =  \frac{(2 \pi)^{3/2}}{2}\int \widetilde{n^{z}}(\bar{k}_{z})
\widetilde{n^{z}}(-\bar{k}_{z}) \widetilde{U^{z}}(\bar{k}_{z}) 
\, d \bar{k}_{z}\;,
\end{eqnarray}
where $\tilde{n}(k)$ is the Fourier transform of the number
density.
In this expression
the axial momentum has been scaled by the laser wavenumber
$\bar{k}_{z} \equiv k_{z}/q$, and the one-dimensional (1D)
axial potential is given 
by 
\begin{equation}
\widetilde{U^{z}}(\bar{k}_{z})=  
 \frac{2}{(2 \pi)^{3/2}} \,
\frac{{\mathrm I} \, \alpha^{2} q^{3} }{8 \pi
\epsilon_{0}^{2} c} \,
Q(\overline{w}_{r}, \bar{k}_{z}) \;,
\label{ueff1}
\end{equation}
with
\begin{eqnarray}
Q(\overline{w}_{r}  , \bar{k}_{z})    &\simeq&   -\frac{2}{3}
\frac{1}{(2 \pi \overline{w}_{r})^{2}}
    + 
2  ( \bar{k}_{z}^{2}-1 )   \exp    \left(
-\frac{(2 \pi \overline{w}_{r})^{2}}{2}	 \right)  \times 
 	 \nonumber \\ 
& & \left\{ \frac{1}{4} (2 \pi \overline{w}_{r})^2 +
 \frac{1}{2}  \exp  \left(\frac{(\bar{k}_{z}^{2}-1 ) (2 \pi
\overline{w}_{r} )^{2}}{2} \right)
  E_{1}  \left(\frac{(\bar{k}_{z}^{2}-1 ) (2 \pi
\overline{w}_{r} )^{2}}{2}  \right)\right.   	  +   \nonumber \\ 
& & \;\;\left.
 \frac{1}{8}  (1- \bar{k}_{z}^{2})(2 \pi \overline{w}_{r})^{4}
 \exp  \left(\frac{(\bar{k}_{z}^{2}-1 ) (2 \pi
\overline{w}_{r} )^{2}}{2} \right)   
E_{1}       \left(\frac{(\bar{k}_{z}^{2}-1 ) (2 \pi
\overline{w}_{r} )^{2}}{2} \right)
   \right\}   \;,
\label{ueff2}
\end{eqnarray}
where $ E_{1}$ is the exponential integral \cite{a+s}.

\begin{figure}[htbp]
\begin{center}
\centerline{\epsfig{figure=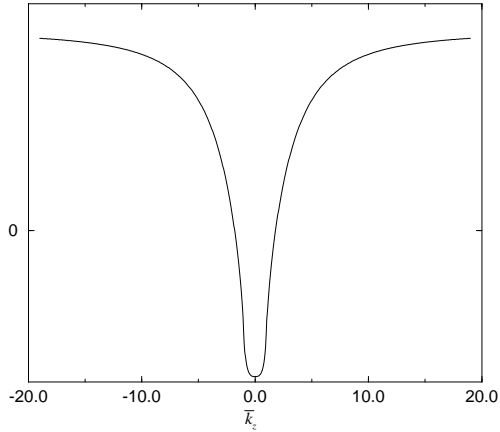,width=8cm}}
\end{center}
\vspace{-4ex}
\caption{The $\bar{k}_{z}$ dependence of the one dimensional axial potential
$ \widetilde{U^{z}}(\bar{k}_{z})$. The radial width has been set at
$\overline{w}_{r}=1/(4 \pi)$.
}
\label{fig:1dpot}
\end{figure}

We plot $\widetilde{U^{z}}(\bar{k}_{z})$ 
in  Fig.~\ref{fig:1dpot}.
It tends to a positive constant value for large $\bar{k}_{z}$,
as verified by the appropriate limit of (\ref{ueff1}) and (\ref{ueff2}):
\begin{eqnarray}
\lim_{\bar{k}_{z} \rightarrow \infty} \widetilde{U^{z}}(\bar{k}_{z})
= \frac{2}{(2 \pi)^{3/2}}
\frac{{\mathrm I} \, \alpha^{2} q^{3} }{8 \pi
\epsilon_{0}^{2} c} \left[ -\frac{2}{3}
\frac{1}{(2 \pi \overline{w}_{r})^{2}} + 
 2  \exp	 \left(
-\frac{(2 \pi \overline{w}_{r})^{2}}{2}	 \right)
\left(\frac{1}{(2 \pi \overline{w}_{r})^{2}} + \frac{1}{2} \right)
\right]\;.
\label{ukzinfty}
\end{eqnarray} 
This means that the axial dipole-dipole potential in coordinate space will
contain a \emph{repulsive delta-function contribution},
which can \emph{stabilize the trapped condensate} 
in addition to  (or instead of)
the short-range s-wave scattering.
Despite this effective short-range repulsion, the overall 
long-range behavior is dictated by the 
$\bar{k}_{z} \rightarrow 0$ limit of $\widetilde{U^{z}}(\bar{k}_{z})$,
which is \emph{attractive}.
When the $\bar{k}_{z}=0$ Fourier component of the total 1D-reduced potential
(the sum of the s-wave scattering pseudo-potential and 
the laser-induced dipole-dipole interaction)
becomes negative, the condensate size decreases drastically until
it becomes \emph{self-trapped within the laser-wavelength}.

\begin{figure}[htbp]
\begin{center}
\centerline{\epsfig{figure=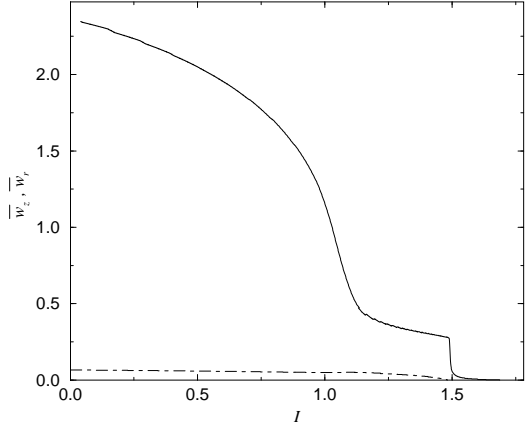,width=8cm}}
\end{center}
\vspace{-4ex}
\caption{Condensate size as a function of the scaled laser
intensity $\mathcal{I}$. Solid line:
$\overline{w}_{z}$, dashed line: $\overline{w}_{r}$.
The trap frequencies have been chosen so that
at  ${\mathcal I}=0$ (i.e.\ without dipole-dipole forces)
the condensate has an aspect ratio of 30:1, with
a radial dimension of $\overline{w}_{r}=1/(4 \pi)$.
}
\label{fig:w30}
\end{figure}

\begin{figure}[htbp]
\begin{center}
\centerline{\epsfig{figure=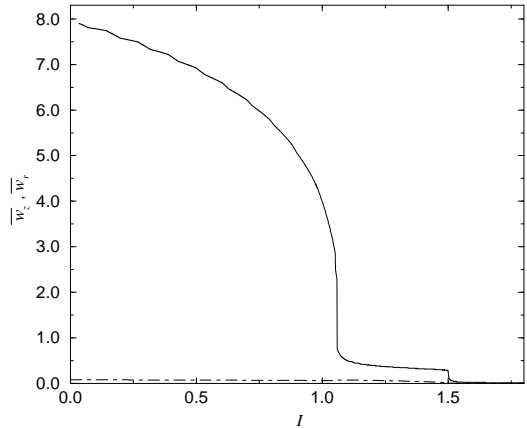,width=8cm}}
\end{center}
\vspace{-4ex}
\caption{Condensate size as a function of the scaled laser
intensity $\mathcal{I}$, as in fig. 3. 
here the trap frequencies are chosen such that
at  ${\mathcal I}=0$ (i.e.\ without dipole-dipole forces)
the condensate has an aspect ratio of 100:1, with
a radial dimension of $\overline{w}_{r}=1/(4 \pi)$.
}
\label{fig:w100}
\end{figure}

From Fig.~\ref{fig:1dpot}, the monotonic form of the axial potential 
suggests that the simplest ansatz for the
axial variational density is also a Gaussian
\begin{equation}
 \widetilde{n^{z}}(\bar{k}_{z}) \equiv \frac{1}{\sqrt{2 \pi}}\exp 
\left[-\frac{(2 \pi \overline{w}_{z})^{2} \bar{k}_{z}^{2}}{4} 
\right].
\end{equation}
Even with this choice, however, the remaining integral in the calculation
of $H_{\mathrm{dd}}$ is difficult
and will here be evaluated numerically.
In the large $N$  limit, the so-called Thomas-Fermi regime, $H_{\mathrm{ke}}$
may be neglected in comparison with the interaction energies and thus the 
energy $H_{{\mathrm TF}}=H_{{\mathrm s}}+H_{{\mathrm dd}}+H_{{\mathrm ho}}$
can be written in the dimensionless form
\begin{eqnarray}
\frac{m }{q^{3} N^{2} \hbar^{2} a}
 H_{{\mathrm TF}}  =   \frac{1}{\sqrt{2 \pi} \; 2 \pi \overline{w}_{z} (2 \pi 
\overline{w}_{r})^{2}}  + 
{\mathcal I} \int \widetilde{n^{z}}^{2}
(\bar{k}_{z})   Q(\overline{w}_{r},\bar{k}_{z}) d \bar{k}_{z} 
   +  \frac{1}{2 N a q} 
\left[\frac{(2 \pi \overline{w}_{r})^{2}}{(q l_{{\mathrm ho}}^{r})^{4}}
+\frac{(2 \pi \overline{w}_{z})^{2}}{2(q l_{{\mathrm ho}}^{z})^{4}} \right]
\;,
\label{eq:HTF}
\end{eqnarray}
where $l_{{\mathrm ho}}^{r}=\sqrt{\hbar/(m \omega_{r})}$ is the harmonic
oscillator length due to a radial trap of frequency $\omega_{r}$,
and similarly for $l_{{\mathrm ho}}^{z}$. Given these trapping
frequencies, the real control parameter that allows one to play with
the ratio of these dipole-dipole forces to the s-wave scattering
is the dimensionless `intensity'
\begin{equation}
{\mathcal I}= \frac{{\mathrm I} \, \alpha^{2} m}{8 \pi \epsilon_{0}^{2} c \hbar^{2} a}.
\end{equation}
A minimization of $ H_{{\mathrm TF}}$ brings out the strong dependence 
of the condensate radius $\overline{w}_{z}$ and 
the weaker dependence of the $\overline{w}_{r}$ radius
upon $\mathcal{I}$ as shown in Figures~\ref{fig:w30} and  \ref{fig:w100}.
For ${\mathcal I} \agt 1.1$ there is a drastic decrease of
$\overline{w}_{z}$
and self-trapping in accordance with the discussion following 
Eq. (\ref{ukzinfty}).
For ${\mathcal I}=3/2$ the system collapses 
and this can be attributed to the instability caused by the static
$r^{-3}$ part of the dipole-dipole potential (\ref{eq:tpot}),
as in \cite{goral}.

To conclude, we have predicted here a new feature of laser-irradiated
cigar-shaped condensates:
their strong compression along the soft axial direction.
In an experiment this compression would provide a
clear signature of laser-induced 
dipole-dipole forces at work.

\end{document}